\documentclass[5p,twocolumn,authoryear]{elsarticle}

\usepackage{amsmath,amssymb,mathtools}
\usepackage{booktabs}
\usepackage{bm}
\usepackage{siunitx}
\usepackage{enumitem}
\usepackage{physics}
\usepackage{footmisc}
\usepackage{xcolor}
\usepackage{hyperref}
\usepackage{float}

\usepackage[utf8]{inputenc}
\DeclareUnicodeCharacter{2212}{-}

\usepackage[T1]{fontenc}
\usepackage{newtxtext,newtxmath}  

\journal{This paper is currently under review by Energy Policy}

\begin{document}
	
	\begin{frontmatter}
	\title{Empirical Validation of a Dual-Defense Mechanism Reshaping Wholesale Electricity Price Dynamics in Singapore} 

		\begin{abstract}
While ex-ante screening and static price caps are global standards for mitigating price volatility, Singapore’s electricity market employs a unique dual-defense mechanism integrating vesting contracts (VC) with a temporary price cap (TPC). Using high-frequency data from 2021 to 2024, this paper evaluates this mechanism and yields three primary findings. First, a structural trade-off exists within the VC framework: while VC quantity (VCQ) suppresses average prices, it paradoxically exacerbates instability via liquidity squeezes. Conversely, VC price (VCP) functions as a tail-risk anchor, dominating at extreme quantiles where VCQ efficacy wanes. Second, a structural break around the 2023 reform reveals a fundamental re-mapping of price dynamics; the previously positive pass-through from offer ratios to clearing prices was largely neutralized post-reform. Furthermore, diagnostics near the TPC threshold show no systematic evidence of strategic bid shading, confirming the TPC’s operational integrity. Third, the dual-defense mechanism exhibits a critical synergy that resolves the volatility trade-off. The TPC reverses the volatility penalty of high VCQ, shifting the elasticity of conditional volatility from a destabilizing 0.636 to a stabilizing -0.213. This synergy enables the framework to enhance tail-risk control while eliminating liquidity-related stability costs. We conclude that this dual-defense mechanism successfully decouples price suppression from liquidity risks, thereby maximizing market stability.
		\end{abstract}

		
		\author[1]{Zhenyu Huang}
		
		\author[1]{Zhao Yuan\corref{cor1}}
		
		\address[1]{Energy Research Institute, Nanyang Technological University, Singapore}

		\cortext[cor1]{Corresponding author, e-mail: zhao.yuan@ntu.edu.sg}

	\end{frontmatter}
	\section{Introduction}
	
	Electricity market liberalization has become the dominant paradigm in global energy reform over the past three decades, aiming to enhance efficiency and reduce costs through competition \cite{joskow2008lessons}. However, many pioneering jurisdictions have encountered challenges such as market power abuse \cite{graf2021market}, complexities arising from the energy transition, and extreme price volatility driven by severe weather events \cite{newbery2002problems}. Consequently, market rules require continuous recalibration. For instance, the California electricity crisis prompted US independent system operators to adopt sophisticated offer caps and automated mitigation procedures \cite{shang2021review}, while the 2021 global energy crisis forced European nations to intervene in energy-only markets (EOM) through temporary revenue caps and questioned the limits of the merit order effect under gas supply shocks \cite{fabra2023reforming, pollitt2024recommendations}. These global experiences, alongside reforms in transitional economies balancing state guidance with market signals and addressing structural issues like overcapacity or cross-border integration \cite{zheng2021between, ding2026electricity, ciarreta2025restructured}, underscore a critical lesson: regulators must continuously seek a balance between fostering competition and curbing market power \cite{fabra2023market, ding2026electricity}.
	
	Unlike jurisdictions that rely heavily on complex ex-ante mitigation procedures, such as the Pennsylvania-New Jersey-Maryland (PJM) interconnection, or those characterized by purely scarcity-driven pricing with high administrative caps, such as the electric reliability council of Texas (ERCOT), Singapore’s liberalization path has retained a distinctive structural feature: the vesting contract (VC). While the national electricity market of Singapore (NEMS) was established in 2003 with a pro-competition mandate \cite{chang2007new}, the regulator deliberately maintained VC not merely as a transitional legacy, but as a mandatory hedge to lock in a portion of generation capacity. Early on, this seemed to follow the standard trajectory of gradual phase-out. \cite{loi2019electricity} noted that the introduction of futures markets in 2015 helped transition the wholesale market into a lower-volatility regime, and \cite{feng2023electricity} observed diminished price outliers as participation grew. Consequently, the energy market authority (EMA) initially planned to roll back VCs, intending to align fully with the standard EOM orthodoxy where market forces decisively allocate resources.
	
	However, the global energy crisis in late 2021 imposed a structural shock on the Singapore market. Triggered by external supply shocks, the volatility of the uniform Singapore energy price (USEP) intensified dramatically. According to the official document released by the EMA \cite{EMA2023}, this extreme volatility forced numerous independent retailers to exit the market due to prohibitive hedging costs, severely disrupting the previously diversified competitive landscape. In response, the regulator adopted a composite regulatory strategy combining structural constraints with dynamic regulation. This approach primarily involved the recalibration of VC, where the previous rollback plan was halted to re-establish VCs as a long-term mechanism covering Non-Contestable Consumer load, thereby structurally weakening generators' incentives for economic withholding by locking in a portion of their revenue ex-ante. Complementing this measure, the temporary price cap (TPC) was introduced in 2023 to serve as a circuit breaker designed to cut off scarcity rents during periods of extreme supply-demand tightness, preventing prices from decoupling from fundamentals.
	
	Despite these policy adjustments, empirical evidence regarding the effectiveness of Singapore's post-crisis regulation mix remains limited. This paper addresses this gap by quantifying how VC \cite{wolak2000empirical} and the TPC shape wholesale pricing outcomes in the NEMS. Using 30-min clearing prices and an integrated mixed-frequency panel spanning 2021 to 2024, we implement a suite of econometric designs, including ordinary least squares (OLS), regime-switching specifications, quantile regression, structural break tests, and exponential generalized autoregressive conditional heteroskedasticity (EGARCH) models, while conditioning on supply-demand fundamentals and fuel cost conditions proxied by futures settlement prices. Our empirical design is guided by prior evidence that forward coverage can discipline spot prices \cite{keles2022forward} while recognizing that contracting choices may be strategic and endogenous \cite{flottmann2025forward}.
	
	Our results suggest that this dual-defense mechanism is associated with improved price stability under crisis conditions. Higher VC levels are linked to a significant reduction in extreme price outcomes, while the TPC is associated with a structural break that neutralizes markup pass-through without inducing systematic pre-trigger braking, even as tail-risk mitigation remains most pronounced when VCQ interacts with the policy regime. These findings provide empirical evidence relevant to Singapore's ongoing market design and offer reference points for other systems seeking to balance efficiency and security during the energy transition.
	
	\section{Literature review}
	
	\subsection{Structural Drivers of Extreme Prices in Singapore’s Energy-Only Market}
	
	In the landscape of global electricity market reform, Singapore adopted an EOM model, similar to the ERCOT and Australia's national electricity market (NEM). This contrasts with markets that utilize Capacity Payments, such as PJM in the United States or the early pool model in the United Kingdom \cite{cicala2022imperfect}. Under the EOM architecture, generation units do not receive guaranteed capacity revenue to cover fixed investment costs. Theoretically, generators must rely largely on energy sales in the spot market to recover fixed and investment costs over the long run. This structural design introduces a fundamental tension: the market must allow for very high scarcity pricing during periods of supply--demand tightness, enabling marginal and peaking units to recover annualized costs within limited operational hours.
	
	For instance, the price cap in ERCOT was historically set as high as $9{,}000 USD/MWh$ \cite{busby2021cascading}, while the ceiling for the USEP is currently set at $4{,}500 SGD/MWh$. However, this high-cap design, while necessary for investment signals, simultaneously creates ample room for extreme price outcomes when the system becomes tight. The absence of long-term capacity revenue also exposes generators to substantial revenue volatility, amplifying incentives for strategic offer behavior such as bidding above marginal cost or near the cap, which may elevate clearing prices during scarcity episodes. Due to the opacity of real-time fuel costs and operational constraints, it is often difficult for regulators to distinguish between legitimate cost pass-throughs and strategic anomalies in bidding curves \cite{de2025investigating}.

	To address vulnerabilities in such settings, international jurisdictions typically rely on distinct regulatory paradigms involving either strict ex-ante screening, such as the three pivotal supplier test in PJM \cite{bowring2013capacity}, or high static hard caps combined with ex-post conduct regulation. However, transplanting these standard mechanisms to Singapore presents a unique structural dilemma. Although the Herfindahl-Hirschman Index has declined following new entries \cite{loi2019electricity}, the NEMS retains inherent oligopolistic characteristics. Implementing strict ex-ante mechanisms would likely bind too frequently in such a concentrated system, hindering price discovery, while relying solely on ex-post punishment is often ineffective due to information asymmetry regarding fuel constraints. Furthermore, reliance on voluntary forward markets to discipline spot prices is constrained by liquidity and institutional frictions. Forward prices often embed risk premia and serve as noisy proxies for expectations rather than sufficient statistics for fundamentals \cite{bonaldo2022relationship, olmstead2025alberta}. Consequently, the market structure lacks the inherent depth required to naturally mitigate extreme price volatility without specific regulation.
	
	This inapplicability of standard global tools necessitates the use of VCs as a structural hedge to mute incentives ex-ante. In this specific regulatory context, VCs function not merely as a transitional instrument but as a durable form of mandatory forward coverage that locks in a share of generator sales and revenue exposure through administered vesting quantities and prices. By reshaping incentives and reducing exposure to extreme spot outcomes, such contracting-based instruments can compress the upper tail of prices during tight conditions. This mechanism aligns with related evidence on how different contracting arrangements affect spot price dynamics \cite{flottmann2024derivatives}.
	
	\subsection{Post-Crisis regulatary Tightening and Rule Recalibration}
	
	The academic consensus on electricity market liberalization has shifted from initial optimism to caution regarding energy-only designs \cite{ema2016review, EMA2023} . Early literature, such as \cite{simshauser2021lessons}, highlighted the efficiency of Australia’s NEM, where the phased removal of VCs \cite{kee2001vesting} allowed competitive price signals to effectively balance supply and demand during stable periods. However, empirical evidence from extreme events suggests that this model can be vulnerable to strategic behavior under tight conditions, for example, \cite{wolak2003diagnosing} analyzed the California electricity market's crisis (2000--2001) and demonstrated that the crisis was not merely due to scarcity, but was significantly exacerbated by suppliers exercising unilateral market power through economic withholding. 
	
	While these episodes underscore vulnerabilities in market design, a comprehensive evaluation by \cite{cicala2022imperfect} on the U.S. electricity sector provides a nuanced counter-perspective. By comparing regions that transitioned to market-based dispatch with those remaining under traditional command-and-control regulation, he finds that market competition significantly improves production efficiency. This gain is primarily driven by enhanced gains from trade and a reduction in out-of-merit costs—the expenses incurred from running uneconomical units when lower-cost alternatives are available. Despite these systemic benefits, the limits of deregulation were further tested by Winter Storm Uri in Texas (2021) and the European energy crunch (2021--2022) \cite{pollitt2024recommendations}. These events, characterized by external fuel-supply shocks and extreme weather, prompted renewed debate on safeguard design and the efficacy of price caps in mitigating volatility without destroying investment incentives.
	
	Singapore's experience during the 2021--2022 energy crisis provides a critical case study paralleling these global stresses. Historically, Singapore followed the liberalization trend by progressively reducing VC levels. However, this reduced structural coverage left the market more exposed. When global gas prices spiked, the resulting volatility in the USEP was not fully explained by cost pass-through alone and placed severe strain on the retail sector. As hedging costs soared and liquidity dried up, the market witnessed the cascading exit of multiple independent retailers. This episode highlighted that in an island system with oligopolistic characteristics, relying solely on market forces without sufficient structural buffering can lead to financial fragility and a deterioration of competitive retail participation.
	
These vulnerabilities prompted a sharp regulatary pivot and a tightening of safeguards. In direct response to the crisis, the EMA implemented the TPC mechanism and effectively halted the planned phase-out of VCs. Despite the significance of this policy recalibration, no existing studies have quantified how the post-crisis combination of retained VCs and the TPC alters wholesale price formation, particularly regarding extreme price outcomes and spikes. More broadly, price dynamics can be shaped by discrete institutional adjustments, macroeconomic shocks, and regime changes beyond standard competitive mechanisms \cite{sirin2023market, sirin2026monetary}, motivating econometric designs that explicitly incorporate policy shifts and state-dependent effects. This study aims to fill that gap.
	

\section{Data and Econometric Strategy}
	
	To empirically evaluate the efficacy of the dual-defense mechanism, we construct a high-frequency dataset spanning from January 1, 2021, to December 31, 2024.
	
	\subsection{Detail and assumption of the Dual-Defense Mechanism}

	Central to our investigation are two regulatory mechanisms that jointly shape Singapore’s wholesale price formation and risk profile. Under the post-2023 framework, the VC is structured into three components with differentiated pricing arrangements. Because the variable-priced component represents a relatively small share of vesting volume, this study does not distinguish among the three VC types in the empirical specifications; instead, we adopt a parsimonious representation that aggregates them into a unified vesting contract price (VCP) and vesting contract quantity (VCQ), and treat the VC as an regulator-mandated price–quantity contract that mechanically reallocates a prescribed share of generation from spot exposure to regulated settlement.

	The TPC functions as a dynamic circuit breaker designed to constrain extreme price outcomes once market conditions enter a sustained stress state. The TPC is governed by the relationship between the moving average price (MAP), defined as a rolling 24-hour average of wholesale prices, and the moving average price threshold (MAPT) \cite{emc_tpc}. Rather than reacting to isolated half-hour spikes, the mechanism is triggered when the MAP exceeds the MAPT, indicating that elevated prices are persistent rather than transitory. Once triggered, the USEP is capped to prevent prolonged decoupling from underlying cost fundamentals and to limit the propagation of high-price episodes for the next 24 hours. 
	
	Conceptually, the VC and the TPC therefore intervene at different layers of the market: the VC reshapes ex ante exposure by contracting a regulated share of volume, whereas the TPC imposes an ex post constraint on the realized price path when system-wide stress pushes the market into a high-price regime.

	\subsection{Data Sources and High-Frequency Reconstruction}

\begin{table*}[ht]
	\centering
	\caption{Data Sources, Variable Definitions, and Vector Construction}
	\label{tab:data_source}
	\footnotesize
	\renewcommand{\arraystretch}{1.3}
	\begin{tabular}{p{0.1\textwidth} p{0.05\textwidth} p{0.25\textwidth} p{0.05\textwidth} p{0.32\textwidth} p{0.05\textwidth}}
		\hline
		\textbf{Category} & \textbf{Variable} & \textbf{Description} & \textbf{Freq.} & \textbf{Processing \& Alignment} & \textbf{Source} \\ \hline
		
		\textbf{Dependent Variables} & $P^{mix}_t$ & Spliced USEP and RUSEP. Proxy for theoretical uncapped cost. & 30-min & Mixed Splicing. & - \\ \cline{2-6}
		& $P^{usep}_t$ & Actual settlement price subject to TPC censoring. & 30-min & Base Frequency. & EMC \\ \cline{2-6}
		& $\hat{h_t}$ & Conditional variance from EGARCH. & 30-min &  Base Frequency.  & - \\ \hline
		
		\textbf{Fundamentals} & $P^{lag}_{t}$ & Lagged USEP & 30-min & Base Frequency. & EMC \\ \cline{2-6}
		($\mathbf{X}$)
		& $D_t$ & System Total Demand & 30-min & Base Frequency (MFMSSA Anchor). & EMC \\ \cline{2-6}
		
		& $SC^*_{t}$ & Refined Supply Cushion & 30-min & MFMSSA Shadow Recovery (Daily $\to$ 30-min). & EMA \\ \cline{2-6}
		
		& $OR^*_{t}$ & Refined Offer Ratio & 30-min & MFMSSA Shadow Recovery (Monthly $\to$ 30-min). & EMA \\ \cline{2-6}
		
		& $F^{lag}_{w,t}$ & Lagged USEP Futures Price & Weekly & ZOH alignment (Weekly $\to$ 30-min). & SGX \\ \cline{2-6}
		
		& $F_{m,t}$ & Fuel Price & Monthly & ZOH alignment (Monthly $\to$ 30-min). & EMA \\ \hline
		
		\textbf{VC Variables} & $VCQ_t$ & Vesting Contract Quantity & 30-min & Base frequency. & EMA \\ \cline{2-6}
		($\mathbf{Y}$) & $VCP_m$ & Vesting Contract  Price & Monthly & Zero-order hold alignment (Monthly $\to$ 30-min). & EMA \\ \hline
		
		\textbf{Other} & $MAP_t$ & Moving Average Price & 30-min & Base frequency. & EMC \\ \cline{2-6}
		\textbf{Variables}& $MAPT_t$ & Moving Average Price Threshold & 30-min & Base frequency. & EMC \\ \hline

		\textbf{Regime} & $R^{pol}$ & Post-Reform Dummy & Event & Binary: 1 if date $\ge$ 2023-07-01; else 0. & - \\ \cline{2-6}
		& $R^{vcp}_t$ & High-Price Regime & 30-min & Binary: 1 if $P^{lag}_{t} > VCP_m$ ; else 0. & - \\ \cline{2-6}
		& $D^{prox}_t$ & MAPT Proximity & 30-min & Binary: 1 if $M_t \le k\%$; else 0.& - \\ \hline
		
		\textbf{Fixed Effects} & $Y$ & Year Dummies & Annual & & - \\ \cline{2-6}
		($\mathbf{Z}$) & $M$ & Month Dummies & Monthly & & - \\ \cline{2-6}
		& $W$ & Weekday Dummies & Daily & & - \\ \hline
	\end{tabular}
\end{table*}

\paragraph{Data description}

The dataset amalgamates market outcomes from the EMC, regulation parameters from the EMA, and futures information from the Singapore Exchange. We anchor the time index to the 30-min clearing interval and obtain a balanced panel with 70,072 observations.

Variables arrive at heterogeneous frequencies. For stepwise administrative and financial series such as $VCP$ and futures prices, we apply zero-order hold (ZOH) alignment. Each observation is carried forward to all subsequent 30-min intervals until the next update, matching the information set available to market participants in real time.

A key challenge concerns supply-side fundamentals. The supply cushion and offer ratio are reported only at daily and monthly frequencies. Direct interpolation would mechanically smooth these series, thereby masking economically significant intra-day market tightness. To preserve the regulator-reported levels while recovering high-frequency scarcity dynamics, we employ mixed-frequency multivariate singular spectrum analysis (MFMSSA) \cite{hassani2019monthly}. Using total system demand as the high-frequency anchor, we extract a synchronized intra-day basis and reconstruct each low-frequency supply variable by projecting its zero-order hold (ZOH) aligned series onto this demand-driven trajectory. This approach ensures the refined series maintain their original low-frequency integrity while restoring the intra-day granularity essential for scarcity and market-power diagnostics.

\paragraph{Regime settings}

We construct three binary indicators to capture state-dependent dynamics:
\begin{enumerate}
	\item High-Price Regime: A state-contingent indicator defined as one when the lagged spot price exceeds the administrative VCP. This identifies periods where the contract strike price becomes economically binding relative to the spot market clearing price.
	\item Policy Regime: A structural break indicator which equals one for all trading periods on or after July 1, 2023, capturing the shift in the regulatory environment following the implementation of the TPC mechanism.
	\item MAPT Proximity Indicator: A conditional dummy variable used to detect strategic avoidance behavior. It takes a value of one when the margin between the rolling cumulative average price and the activation threshold falls within a specified sensitivity window (e.g., 5\%, 10\%, or 15\%).
\end{enumerate}

\paragraph{Notation and coefficient convention}

Table~\ref{tab:data_source} also defines the construction of the regressor blocks.
We partition the covariates into three vectors: (i) $\mathbf{X}_t$ collects core fundamentals, (ii)$\mathbf{Y}_t$ collects VC-related variables, and (iii) $\mathbf{Z}_t$ collects fixed effects.
These blocks are mapped to coefficient vectors $\boldsymbol{\beta}$, $\boldsymbol{\delta}$, and $\boldsymbol{\gamma}$, respectively.
The intercept is denoted by $\alpha$, and $\varepsilon_t$ denotes the residual of the mean-equation innovation.

For any regime indicator $R^{\cdot}_t$, we adopt a unified naming rule: coefficients on $\mathbf{Y}_t$ are decomposed into a baseline component $\boldsymbol{\gamma}_{base}$ and an incremental regime component $\boldsymbol{\gamma}_{reg}$, so that the implied slope on $\mathbf{Y}_t$ equals $\boldsymbol{\gamma}_{base}$ when $R^{\cdot}_t=0$ and $\boldsymbol{\gamma}_{base}+\boldsymbol{\gamma}_{reg}$ when $R^{\cdot}_t=1$.
The regime intercept shift is captured by $\theta$, the coefficient on $R^{\cdot}_t$ itself.

\subsection{Quantifying the Stabilizing Efficacy of VCs}
\label{sec:vc_analysis}

To assess the effectiveness of the VC mechanism, we employ the mix price $P^{mix}_t$ as the dependent variable to eliminate the influence of TPC in section \ref{sec:vc_analysis}.

\subsubsection{Baseline and Regime-Switching Mean Specifications}

We begin by establishing the fundamental elasticity of spot prices to vesting VCQ. The baseline specification identifies the average treatment effect:

\begin{equation}
	\ln P^{mix}_t = \alpha + \mathbf{X}_t \boldsymbol{\beta} + \mathbf{Z}_t \boldsymbol{\delta} + \mathbf{Y}_t \boldsymbol{\gamma} + \varepsilon_t
	\label{eq:baseline_mean}
\end{equation}

To test the hypothesis that VCs function primarily as a crisis dampener, we employ a regime-switching design:

\begin{equation}
	\ln P^{mix}_t = \alpha + \mathbf{X}_t \boldsymbol{\beta} + \mathbf{Z}_t \boldsymbol{\delta} + \mathbf{Y}_t \boldsymbol{\gamma}_{base} + \theta R^{vcp}_t + R^{vcp}_t \cdot \mathbf{Y}_t \boldsymbol{\gamma}_{reg} + \varepsilon_t
	\label{eq:regime_mean}
\end{equation}


\subsubsection{Conditional Quantile Specification}

Mean regressions may mask heterogeneous effects in the upper tail where price spikes concentrate. We estimate the conditional quantile regression model:

\begin{equation}
	Q_{\tau}\!\left(\ln P^{mix}_t \mid \mathbf{X}_t,\mathbf{Z}_t,\mathbf{Y}_t\right)
	= \alpha(\tau) + \mathbf{X}_t \boldsymbol{\beta}(\tau)
	+ \mathbf{Z}_t \boldsymbol{\delta}(\tau)
	+ \mathbf{Y}_t \boldsymbol{\gamma}(\tau)
	\label{eq:quantile_mean}
\end{equation}
where $\tau \in (0,1)$ denotes the quantile level.
The coefficient vector $\boldsymbol{\gamma}(\tau)$ measures how VC shifts the $\tau$-th conditional quantile of the price distribution.
If the magnitude of $\boldsymbol{\gamma}(\tau)$ increases toward the upper tail, it indicates that VCs exert a stronger dampening effect precisely during spike-prone conditions, consistent with a tail-truncation role.

\subsubsection{Impact on Volatility Dynamics}
\label{sec:garch_model}

To quantify how the VC mechanism reshapes market turbulence, we adopt a two-stage approach that accounts for the asymmetric nature of price volatility and its structural persistence.

Step 1: Volatility Extraction via EGARCH. First, we isolate the idiosyncratic price innovations $\varepsilon_t$ from the conditional mean. Distinct from the standard GARCH framework, we employ an EGARCH specification to model the conditional variance. This approach models the logarithm of the conditional variance, $\ln h_t$, which ensures that the estimated volatility remains strictly positive without imposing parameter constraints and allows for asymmetric responses to positive and negative shocks. The mean equation incorporates the full set of fundamental vectors $\mathbf{X}_t$, policy variables $\mathbf{Y}_t$, and fixed effects $\mathbf{Z}_t$ to ensure the residuals are orthogonal to observable drivers. The conditional variance is governed by the following process:

\begin{equation}
	\label{eq:egarch_generation}
	\ln h_t = \omega + \beta \ln h_{t-1} + \gamma \frac{\varepsilon_{t-1}}{\sqrt{h_{t-1}}} + \alpha \left( \left| \frac{\varepsilon_{t-1}}{\sqrt{h_{t-1}}} \right| - \sqrt{\frac{2}{\pi}} \right)
\end{equation}

In this framework, $\omega$ is the intercept, and $\beta$ captures the persistence of log-volatility. The term $\gamma$ captures the sign effect, testing whether positive and negative price shocks have differential impacts on future volatility, while $\alpha$ measures the magnitude effect of the standardized shock. The parameter $\sqrt{2/\pi}$ represents the expected value of the absolute standardized residual under the normality assumption. By estimating this system, we extract the conditional variance series $\hat{h}_t$ to serve as the dependent variable for the subsequent structural analysis.

Step 2: Dynamic Volatility Regression. In the second stage, we investigate the determinants of the estimated volatility. We treat the logarithm of the extracted conditional variance, $\ln \hat{h}_t$, as the dependent variable. To explicitly account for the strong persistence characteristic of high-frequency volatility series, we augment the structural regression with a lagged dependent variable $\ln \hat{h}_{t-1}$. The specification is defined as:

\begin{equation}
	\label{eq:vol_regression}
	\begin{split}
		\ln \hat{h}_t = & \alpha + \rho \ln \hat{h}_{t-1}
		+ \mathbf{X}_t \boldsymbol{\beta}
		+ \mathbf{Y}_t \boldsymbol{\gamma}_{base}
		+ \mathbf{Z}_t \boldsymbol{\delta}  \\
		& + \theta R^{vcp}_t + R^{vcp}_t \cdot \mathbf{Y}_t \boldsymbol{\gamma}_{reg} + \eta_t
	\end{split}
\end{equation}

where $\rho$ measures the autoregressive persistence of volatility after controlling for structural covariates. The coefficient vectors $\boldsymbol{\gamma}_{base}$ and $\boldsymbol{\gamma}_{reg}$ capture the baseline and regime-dependent impacts of the vesting mechanism on market stability, respectively. The term $\eta_t$ represents the error term of the second-stage regression. This dynamic specification ensures that the estimated coefficients on the policy variables reflect their net contribution to volatility formation, conditional on the inherent inertia of the variance process.

\subsection{Structural Breaks and Threshold Proximity}
\label{sec:method_tpc}

We employ two complementary identification strategies to isolate the structural impact of the TPC on strategic bidding behavior.

\subsubsection{Regime-Switching Structural Break Model}

As a complement to the regime-based analysis, we exploit the policy implementation date as an explicit time-structure break. The TPC reform introduced on 2023-07-01 constitutes a structural change in the price formation mechanism because it targets the clearing/settlement price directly. Even when the market is not in a high-price activation state, the presence of a binding cap rule can alter bidding incentives by changing the expected payoff to aggressive offers and by compressing the price response to markup-related signals. Accordingly, this subsection tests whether the reform attenuates the pass-through from observed bidding and markup proxies to $P^{usep}_t$. We implement an elasticity-shift specification that allows the slopes on fundamental markup signals to change discretely after the policy date, isolating the deterrence channel of the TPC from contemporaneous demand–supply fundamentals.
\begin{equation}
	\begin{split}
		\ln P^{usep}_t
		=\, & \alpha
		+ \mathbf{X}_t \boldsymbol{\beta}_{base}
		+ \mathbf{Y}_t \boldsymbol{\gamma}
		+ \mathbf{Z}_t \boldsymbol{\delta}
		+ \varepsilon_t\\
		& + \theta R^{pol}_t
		+ R^{pol}_t \cdot \mathbf{X}_t \boldsymbol{\beta}_{pol}
	\end{split}
\end{equation}

\subsubsection{Threshold Proximity and Avoidance Model}

As the market approaches the TPC activation threshold, the payoff structure facing bidders changes discretely. This is because a marginal increase in the MAP can switch the system from the standard cap of 4,500 SGD/MWh to a significantly lower long-run price threshold prescribed by the ISO. This drastic reduction in the effective price ceiling creates incentives for strategic price adjustment as the trigger condition nears. In particular, participants may moderate markups or reallocate bidding pressure to delay activation and avoid the more restrictive price regime. Such behaviors are inherently local and may not be well captured by average elasticities estimated over the full sample.

To examine behavior near the activation threshold, we estimate a proximity-based model:
\begin{equation}
	\begin{split}
		\ln P^{usep}_t
		=\, & \alpha
		+ \mathbf{X}_t \boldsymbol{\beta}
		+ \mathbf{Y}_t \boldsymbol{\gamma}
		+ \mathbf{Z}_t \boldsymbol{\delta}\\
		& + \lambda_1 D^{prox}_t
		+ \lambda_2 D^{prox}_t \cdot M_t
		+ \varepsilon_t
	\end{split}
\end{equation}
where $\lambda_1$ and $\lambda_2$ are coefficients; $M_t$ is the remaining margin to the trigger level, defined as $M_t = MAPT_t - MAP_t$; $D^{prox}_t$ is a proximity indicator that equals 1 when $M_t$ lies within the bottom $x\%$ of its empirical distribution and 0 otherwise.

\subsection{Modeling Policy Synergy and Volatility}
\label{sec:method_net_effect}

Finally, we evaluate the aggregate impact of the dual-defense mechanism using the realized settlement price $P^{usep}_t$. Relative to the VC-only analysis based on the counterfactual mix price $P^{mix}_t$, this section re-estimates the same empirical framework under the capped price environment and tests policy complementarity: whether the introduction of the TPC alters the stabilizing elasticity of VC with respect to $P^{usep}_t$.

\subsubsection{Policy Synergy Identification}

Building on the baseline mean specification, we introduce post-policy interactions on the VC block to identify whether the VC elasticity shifts after the TPC reform:

\begin{equation}
	\begin{split}
		\ln P^{usep}_t
		=\, & \alpha
		+ \mathbf{X}_t \boldsymbol{\beta}
		+ \mathbf{Y}_t \boldsymbol{\gamma}_{base}
		+\mathbf{Z}_t \boldsymbol{\delta}\\
		& + \theta R^{pol}_t
		+ R^{pol}_t \cdot \mathbf{Y}_t \boldsymbol{\gamma}_{pol}
		+ \varepsilon_t
	\end{split}
	\label{eq:synergy_mean}
\end{equation}

This specification isolates a discrete post-policy shift in the VC elasticity under the capped settlement price. The complementarity effect is summarized by $\boldsymbol{\gamma}_{pol}$, which captures how the pass-through from VC to $P^{usep}_t$ changes after the reform.

\subsubsection{Distributional Synergy and Tail Risk}

To assess whether the complementarity is concentrated in spike-prone conditions, we extend the conditional quantile specification with the same policy interactions:

\begin{equation}
	\begin{split}
	Q_{\tau}\!\left(\ln P^{usep}_t \mid \mathcal{F}_{t-1}\right)
	=& \alpha(\tau)
	+ \mathbf{X}_t \boldsymbol{\beta}(\tau)
	+ \mathbf{Y}_t \boldsymbol{\gamma}_{base}(\tau)\\
	&+ \mathbf{Z}_t \boldsymbol{\delta}(\tau) 
	+ \theta(\tau) R^{pol}_t \\
	&+ R^{pol}_t \cdot \mathbf{Y}_t \boldsymbol{\gamma}_{pol}(\tau)
	\end{split}
	\label{eq:synergy_qr}
\end{equation}

We track $\boldsymbol{\gamma}_{pol}(\tau)$ across $\tau$. A strengthening toward upper quantiles indicates that the policy interaction primarily operates in the tail, consistent with the TPC reinforcing the VC stabilizing channel during extreme price conditions rather than uniformly across the distribution.

\subsubsection{Volatility Dampening Analysis}

Finally, we examine whether the dual-defense mechanism alters volatility transmission under the capped price environment. Adhering to the two-stage procedure outlined in Section~\ref{sec:garch_model}, we first extract the conditional variance series $\hat{h}_t$ via the EGARCH specification applied to the residuals derived from Eq. (\ref{eq:synergy_mean}). Subsequently, we estimate the second-stage dynamic volatility regression. To account for volatility persistence, we augment the specification with the lagged dependent variable:

\begin{equation}
	\label{eq:synergy_vol_regression}
	\begin{split}
		\ln \hat{h}_t
		=\, & \alpha + \rho \ln \hat{h}_{t-1}
		+ \mathbf{X}_t \boldsymbol{\beta}
		+ \mathbf{Y}_t \boldsymbol{\gamma}_{base}
		+ \mathbf{Z}_t \boldsymbol{\delta} \\
		& + \theta R^{pol}_t
		+ R^{pol}_t \cdot \mathbf{Y}_t \boldsymbol{\gamma}_{pol}
		+ \eta_t
	\end{split}
\end{equation}

The interaction coefficients $\boldsymbol{\gamma}_{pol}$ capture the post-policy structural shift in the volatility impact of VC conditional on the capped settlement price environment.

\section{Empirical Results}

\subsection{Structural Stabilizer Effectiveness of the VC Mechanism}
\label{sec:vc_effectiveness}

This section evaluates the stabilizing role of the VC mechanism in the absence of price censoring, using the composite wholesale price $P^{mix}_t$.

\subsubsection{Regime-Dependent Price Formation under the VC Framework}

\begin{table}[ht]
	\centering
	\caption{Baseline and Regime-Switching OLS Estimation Results}
	\label{tab:vc_regime_ols}
	\footnotesize
	\renewcommand{\arraystretch}{1.2}
	\setlength{\tabcolsep}{6pt}
	\begin{tabular}{lcc}
		\toprule
		\textbf{Dependent variable: $\ln P^{mix}_t$} & \textbf{Baseline OLS} & \textbf{Regime-Switching OLS} \\
		\midrule
		\multicolumn{3}{l}{\textbf{Panel A: VC Variables}} \\
		\textbf{$\ln VCQ_t$} & -0.1026$^{***}$ & -0.0591$^{***}$ \\
		& (-13.21) & (-7.97) \\
		\textbf{$\ln VCP_m$} & -0.0563$^{***}$ & -0.0199 \\
		& (-4.61) & (-1.46) \\
		\midrule
		\multicolumn{3}{l}{\textbf{Panel B: Regime Interactions}} \\
		\textbf{$R^{vcp}_t$} &  & 1.2183$^{***}$ \\
		&  & (7.89) \\
		\textbf{$R^{vcp}_t \times \ln VCQ_t$} &  & -0.4026$^{***}$ \\
		&  & (-12.45) \\
		\textbf{$R^{vcp}_t \times \ln VCP_m$} &  & -0.0672$^{***}$ \\
		&  & (-2.83) \\
		\midrule
		\multicolumn{3}{l}{\textbf{Panel C: Core Fundamentals}} \\
		\textbf{$\ln P^{lag}_{t}$} & 0.8571$^{***}$ & 0.8350$^{***}$ \\
		& (170.36) & (116.00) \\
		\textbf{$\ln D_t$} & 0.4325$^{***}$ & 0.3900$^{***}$ \\
		& (25.39) & (25.14) \\
		\textbf{$\ln F_{m,t}$} & -0.0224$^{*}$ & -0.0060 \\
		& (-2.24) & (-0.58) \\
		\textbf{$\ln F^{lag}_{w,t}$} & 0.0642$^{***}$ & 0.0654$^{***}$ \\
		& (9.53) & (9.65) \\
		\textbf{$\ln SC^*_t$} & -0.0493$^{***}$ & -0.0481$^{***}$ \\
		& (-8.56) & (-8.24) \\
		\textbf{$\ln OR^*_t$} & 0.0752$^{***}$ & 0.0668$^{***}$ \\
		& (7.96) & (7.07) \\
		\midrule
		\multicolumn{3}{l}{\textbf{Panel D: Model Diagnostics}} \\
		$R^2$ & 0.900 & 0.901 \\
		Adj.\ $R^2$ & 0.900 & 0.901 \\
		AIC & -27,900 & -28,450 \\
		BIC & -27,630 & -28,150 \\
		\bottomrule
		\multicolumn{3}{p{0.92\linewidth}}{\scriptsize
			\textbf{Note:} t-statistics in parentheses. $^{***}$, $^{**}$, and $^{*}$ denote significance at the 1\%, 5\%, and 10\% levels, respectively. Standard errors are heteroscedasticity and autocorrelation (HAC) robust using 48 lags. All specifications include month, year, and weekday fixed effects.}
	\end{tabular}
\end{table}

Table \ref{tab:vc_regime_ols} reports the estimation results for the baseline log-linear specification and the regime-dependent interaction model. Allowing VC effects to vary when the VCP threshold is breached improves model fit in terms of information criteria. The Akaike information criterion (AIC) decreases from -27,900 in the baseline specification to -28,450 in the regime-dependent specification, while the $R^2$ remains essentially unchanged at 0.901. This pattern indicates that the improvement is driven by a better characterization of regime-dependent slope changes around the vesting threshold rather than a mechanical increase in explained variance.

The baseline model implies a stabilizing effect of VCQ, with an elasticity of -0.1026. The regime-dependent model shows that this effect differs systematically across price regimes. In the normal regime, the elasticity with respect to contract quantity is -0.0591. When the market enters the high-price regime, the additional interaction effect of -0.4026 becomes active, so the implied elasticity in the high-price regime is -0.4617. In absolute value, this corresponds to 7.81 times the normal-regime elasticity, indicating that vesting quantities play a markedly stronger price-dampening role during threshold-breaching scarcity episodes.

The VCP exhibits a more conditional role. In the regime-dependent specification, the baseline coefficient on the VCP is not statistically distinguishable from zero, indicating that when spot prices remain below the vesting benchmark, variation in the VCP level does not materially constrain spot outcomes. In contrast, the interaction term is negative and statistically significant, implying that the VCP becomes relevant mainly in threshold-breaching periods. The combined effect in the high-price regime is therefore -0.0871, indicating additional downward pressure on prices precisely when the threshold is breached.

The regime indicator is positive and statistically significant, capturing a scarcity-related price component associated with threshold-breaching periods that is not explained by observed fundamentals. Among the controls, the contemporaneous fuel price is significant in the baseline model but becomes insignificant in the regime-dependent specification. This is consistent with fuel price being measured at a monthly frequency and partially absorbed by the lagged forward-price proxy, which captures higher-frequency cost expectations and market conditions more directly. Price persistence remains strong across specifications, with the lagged price coefficient staying close to unity, reflecting the pronounced autoregressive structure of wholesale electricity prices.


\begin{table*}[ht]
	\centering
	\caption{Quantile Regression Estimates Across the Price Distribution}
	\label{tab:quantile_results_mix}
	\footnotesize
	\renewcommand{\arraystretch}{1.15}
	\setlength{\tabcolsep}{6pt}
	\begin{tabular}{l p{2cm} p{2cm} p{2cm} p{2cm} p{2cm} p{2cm} }
		\toprule
		\textbf{Dependent variable: $\ln P^{mix}_t$} & \textbf{Q50} & \textbf{Q75} & \textbf{Q90} & \textbf{Q95} & \textbf{Q97.5} & \textbf{Q99} \\
		\midrule
		\multicolumn{7}{l}{\textbf{Panel A: VC Variables}} \\
		\textbf{$\ln VCQ_t$}  & -0.0113$^{***}$ & -0.0538$^{***}$ & -0.0654$^{***}$ & -0.0710$^{***}$ & -0.0479$^{**}$ & -0.0811$^{**}$ \\
		& (-14.25) & (-12.87) & (-7.10) & (-4.92) & (-2.16) & (-2.08) \\
		\textbf{$\ln VCP_m$}  & -0.0071$^{***}$ & -0.0450$^{***}$ & -0.0778$^{***}$ & -0.1866$^{***}$ & -0.2385$^{***}$ & -0.2430$^{***}$ \\
		& (-6.11) & (-7.23) & (-5.67) & (-8.70) & (-7.25) & (-4.21) \\
		\midrule
		\multicolumn{7}{l}{\textbf{Panel B: Core Fundamentals}} \\
		\textbf{$\ln P^{lag}_{t}$}   & 0.9876$^{***}$ & 0.9678$^{***}$ & 0.9734$^{***}$ & 1.0027$^{***}$ & 1.0703$^{***}$ & 1.1874$^{***}$ \\
		& (5068.84) & (893.21) & (432.21) & (288.23) & (201.80) & (120.86) \\
		\textbf{$\ln D_t$}        & 0.0345$^{***}$ & 0.1625$^{***}$ & 0.3613$^{***}$ & 0.5095$^{***}$ & 0.7059$^{***}$ & 0.9966$^{***}$ \\
		& (29.60) & (26.33) & (27.04) & (24.43) & (21.69) & (16.96) \\
		\textbf{$\ln F_{m,t}$}    & -0.0008 & -0.0167$^{***}$ & 0.0161 & 0.0622$^{***}$ & 0.0766$^{***}$ & 0.1300$^{***}$ \\
		& (-0.89) & (-3.66) & (1.58) & (3.87) & (3.04) & (2.94) \\
		\textbf{$\ln F^{lag}_{w,t}$} & 0.0031$^{***}$ & 0.0510$^{***}$ & 0.1250$^{***}$ & 0.1822$^{***}$ & 0.2212$^{***}$ & 0.3060$^{***}$ \\
		& (8.28) & (25.59) & (28.36) & (26.66) & (21.17) & (16.44) \\
		\textbf{$\ln SC^*_t$}     & -0.0034$^{***}$ & -0.0262$^{***}$ & -0.0827$^{***}$ & -0.1290$^{***}$ & -0.1773$^{***}$ & -0.2416$^{***}$ \\
		& (-12.62) & (-19.04) & (-27.72) & (-27.37) & (-24.37) & (-17.27) \\
		\textbf{$\ln OR^*_t$}     & 0.0062$^{***}$ & 0.0219$^{***}$ & 0.0552$^{***}$ & 0.0978$^{***}$ & 0.1607$^{***}$ & 0.2084$^{***}$ \\
		& (9.24) & (6.16) & (7.11) & (8.24) & (8.80) & (6.74) \\
		\midrule
		\multicolumn{7}{l}{\textbf{Panel C: Model Diagnostics}} \\
		Pseudo $R^2$ & 0.7958 & 0.7967 & 0.7592 & 0.7370 & 0.7243 & 0.6899 \\
		\bottomrule
		\multicolumn{7}{p{0.95\linewidth}}{\scriptsize
			\textbf{Note:} t-statistics are reported in parentheses. $^{***}$, $^{**}$, and $^{*}$ denote significance at the 1\%, 5\%, and 10\% levels, respectively. All specifications include month, year, and weekday fixed effects. Pseudo $R^2$ measuring the local goodness-of-fit at each specific quantile by comparing the sum of weighted absolute deviations of the full model against a restricted intercept-only model.}
	\end{tabular}
\end{table*}

\subsubsection{Tail Effects and State Dependence of VC}
\label{sec:quantile_reg}

Average regressions summarize a single conditional mean response and can therefore understate how vesting parameters operate across market states. Quantile regressions provide a more direct view of state dependence by tracing how the same set of fundamentals and vesting variables map into different parts of the conditional price distribution under the log specification.

The results point to a clear asymmetry between typical conditions and scarcity conditions. Around the center of the distribution, vesting effects are statistically significant but economically small. At the median, the elasticity with respect to VCQ is $-0.0113$ and the elasticity with respect to the VCP is $-0.0071$. These magnitudes suggest that when prices lie near the middle of the conditional distribution, marginal price formation is dominated by routine fundamentals, with vesting parameters exerting only limited quantitative influence.

This picture changes in the upper tail, where tight conditions make the vesting mechanism more consequential. The elasticity of VCQ becomes more negative as one moves up the distribution, taking values of $-0.0538$ at the 75th percentile, $-0.0654$ at the 90th percentile, and $-0.0710$ at the 95th percentile. The monotonic strengthening of the negative effect is consistent with quantity coverage exerting stronger price-dampening pressure when the market tightens, because larger contracted volume reduces the exposure of the residual spot position to marginal imbalances.

The VCP exhibits a distinct pattern concentrated on extreme outcomes. Its elasticity is moderate through the upper-middle of the distribution, at $-0.0450$ in the 75th percentile and $-0.0778$ in the 90th percentile, but it steepens sharply in the extreme tail. At the 95th percentile the estimate is $-0.1866$, and it reaches $-0.2385$ at the 97.5th percentile, remaining sizeable at $-0.2430$ at the 99th percentile. This gradient is consistent with a mechanism in which the VCP has limited bite in routine states but becomes increasingly important as an effective reference that constrains tail realizations.

The behavior of tightness indicators supports the same scarcity interpretation. The supply cushion elasticity becomes progressively more negative as one moves up the distribution, from $-0.0034$ at the median to $-0.2416$ at the 99th percentile. The offer ratio elasticity rises monotonically over the same range, from $0.0062$ at the median to $0.2084$ at the 99th percentile. These patterns indicate that tail prices are disproportionately shaped by tightening physical conditions and offer-side behavior—precisely the environment in which vesting parameters are expected to matter most.

Taken together, the quantile evidence supports a state-dependent interpretation of the vesting mechanism under the log specification. Vesting quantity is associated with broader price-dampening effects that become economically meaningful when prices are high, while the VCP primarily constrains the extreme upper tail. The overall implication is that the mechanism mitigates tail risk while remaining comparatively neutral in magnitude around typical price conditions.

\subsubsection{Volatility Dynamics and the Liquidity Trade-off}
\label{sec:volatility_analysis}

\begin{table}[ht]
	\centering
	\caption{Impact of VCs on Conditional Volatility (EGARCH Model)}
	\label{tab:garch_volatility}
	\footnotesize
	\renewcommand{\arraystretch}{1.15}
	\setlength{\tabcolsep}{6pt}
	\begin{tabular}{lcc}
		\toprule
		\textbf{Dependent variable: $\ln \hat{h}_t$} & \textbf{Baseline} & \textbf{Regime-Switching} \\
		\midrule
		\multicolumn{3}{l}{\textbf{Panel A: VC Variables}} \\
		\textbf{$\ln VCQ_t$} & 0.0005 & -0.0423$^{**}$ \\
		& (0.028) & (-2.565) \\
		\textbf{$\ln VCP_m$} & 0.0134 & 0.0577$^{*}$ \\
		& (0.481) & (1.844) \\
		\midrule
		\multicolumn{3}{l}{\textbf{Panel B: Regime Interactions}} \\
		\textbf{$R^{vcp}_t$} &  & -0.4831 \\
		&  & (-1.449) \\
		\textbf{$R^{vcp}_t \times \ln VCQ_t$} &  & 0.6072$^{***}$ \\
		&  & (8.688) \\
		\textbf{$R^{vcp}_t \times \ln VCP_m$} &  & -0.1386$^{***}$ \\
		&  & (-2.774) \\
		\midrule
		\multicolumn{3}{l}{\textbf{Panel C: Core Fundamentals}} \\
		\textbf{$\ln P^{lag}_{t}$} & 0.3546$^{***}$ & 0.3533$^{***}$ \\
		& (23.417) & (18.061) \\
		\textbf{$\ln D_t$} & -0.0440 & 0.0147 \\
		& (-1.318) & (0.472) \\
		\textbf{$\ln F_{m,t}$} & -0.0144 & -0.0296 \\
		& (-0.673) & (-1.364) \\
		\textbf{$\ln F^{lag}_{w,t}$} & -0.0064 & -0.0198 \\
		& (-0.483) & (-1.439) \\
		\textbf{$\ln SC^*_t$} & -0.0090 & -0.0132$^{*}$ \\
		& (-1.314) & (-1.918) \\
		\textbf{$\ln OR^*_t$} & 0.0508$^{**}$ & 0.0499$^{**}$ \\
		& (2.494) & (2.433) \\
		\midrule
		\multicolumn{3}{l}{\textbf{Panel D: Model Diagnostics}} \\
		$R^2$ & 0.934 & 0.934 \\
		\bottomrule
		\multicolumn{3}{p{0.95\linewidth}}{\scriptsize
			\textbf{Note:} t-statistics in parentheses. $^{***}$, $^{**}$, and $^{*}$ denote significance at the 1\%, 5\%, and 10\% levels, respectively. The dependent variable is the log conditional variance $\ln \hat{h}_t$ extracted from an EGARCH(1,1) specification. The ARCH-LM test ($p>0.05$) indicates no remaining ARCH effects in the standardized residuals. All specifications include temporal fixed effects.}
	\end{tabular}
\end{table}

Table \ref{tab:garch_volatility} reports the second-stage volatility regressions, where the dependent variable is the logarithm of the conditional variance inferred from the first-stage EGARCH price equation. Prior to interpreting the coefficients, the validity of the two-stage approach is formally assessed. As indicated by the diagnostic tests, the baseline OLS specification exhibits severe conditional heteroscedasticity, strongly rejecting the null hypothesis of homoscedastic errors. In contrast, after filtering volatility dynamics through the EGARCH specification, the ARCH-LM test applied to the standardized residuals yields a p-value of 0.091, implying a failure to reject the null hypothesis of no remaining ARCH effects. This result confirms that the EGARCH filter effectively captures volatility clustering and validates the subsequent analysis of volatility determinants.

Turning to the effects of VCs, the results reveal a clearly state-dependent pattern. In the baseline specification, the coefficient on VCQ is statistically indistinguishable from zero, indicating that VCQ does not exert a systematic influence on conditional variance under normal market conditions. Once regime interactions are introduced, the average effect of VCQ becomes negative and statistically significant, suggesting that higher vesting quantities are, on average, associated with lower volatility outside scarcity episodes.

This stabilizing effect reverses sharply when the price cap regime is activated. The interaction between the high-price regime indicator and VCQ is positive and highly significant, with an estimated coefficient of 0.6072. Combined with the baseline VCQ effect in the interaction model, this implies a substantially higher elasticity of conditional variance with respect to VCQ during threshold-breaching periods. The evidence indicates that reductions in residual spot-market depth become destabilizing precisely when the market is tight, amplifying the volatility of price innovations during scarcity episodes.

VCP play a distinct and asymmetric role. In the baseline volatility regression, the coefficient on VCP is statistically insignificant. In the normal regime of the interaction model, it enters with a weakly positive coefficient. By contrast, the interaction between VCP and the high-price regime is negative and statistically significant, with an estimated coefficient of −0.1386. This pattern suggests that VCP becomes highly relevant for volatility during threshold-breaching periods, where it provides an effective anchoring mechanism that limits the amplitude of price fluctuations.

Among the control variables, lagged prices exert a strong and highly significant influence on conditional variance, reflecting pronounced volatility persistence. The supply cushion enters with a negative coefficient and is weakly significant in the regime-dependent specification, indicating that larger capacity buffers help suppress volatility. Notably, the offer ratio displays a significant positive association with volatility, suggesting that aggressive bidding strategies contribute to price instability. Other fundamentals, including demand and fuel prices, display limited explanatory power once volatility dynamics and institutional interactions are properly accounted for.

\subsection{Strategic Deterrence and Behavioral Shifts under the TPC Mechanism}

\subsubsection{Structural Breaks and Strategic Deterrence}
\label{sec:structural_breaks}

\begin{table}[ht]
	\centering
	\caption{Structural Break and Deterrence Effect Estimation Results}
	\label{tab:deterrence_results}
	\footnotesize
	\renewcommand{\arraystretch}{1.2}
	\setlength{\tabcolsep}{6pt}
	\begin{tabular}{lcc}
		\toprule
		\textbf{Dependent variable: $\ln P^{usep}_t$} & \textbf{Baseline} & \textbf{Structural Break} \\
		\midrule
		\multicolumn{3}{l}{\textbf{Panel A: Behavioral Determinants}} \\
		\textbf{$\ln SC^*_t$} & -0.0489$^{***}$ & -0.0985$^{***}$ \\
		& (-8.54) & (-12.00) \\
		\textbf{$\ln OR^*_t$} & 0.0756$^{***}$ & 0.1167$^{***}$ \\
		& (8.00) & (9.46) \\
		\midrule
		\multicolumn{3}{l}{\textbf{Panel B: Regime Interactions}} \\
		\textbf{$R^{pol}_t$} &  & -0.0724$^{***}$ \\
		&  & (-7.68) \\
		\textbf{$R^{pol}_t \times \ln OR^*_t$} &  & -0.1405$^{***}$ \\
		&  & (-7.53) \\
		\textbf{$R^{pol}_t \times \ln SC^*_t$} &  & 0.0673$^{***}$ \\
		&  & (7.64) \\
		\midrule
		\multicolumn{3}{l}{\textbf{Panel C: Controls}} \\
		\textbf{$\ln P^{lag}_{t}$} & 0.8565$^{***}$ & 0.8502$^{***}$ \\
		& (169.44) & (166.36) \\
		\textbf{$\ln D_t$} & 0.4325$^{***}$ & 0.4160$^{***}$ \\
		& (25.40) & (24.89) \\
		\textbf{$\ln F_{m,t}$} & -0.0232 & 0.0010 \\
		& (-2.34) & (0.06) \\
		\textbf{$\ln F^{lag}_{w,t}$} & 0.0646$^{***}$ & 0.0424$^{***}$ \\
		& (9.56) & (6.31) \\
		\textbf{$\ln VCQ_t$} & -0.1022$^{***}$ & -0.1604$^{***}$ \\
		& (-13.29) & (-17.30) \\
		\textbf{$\ln VCP_m$} & -0.0561$^{***}$ & -0.0170 \\
		& (-4.63) & (-1.28) \\
		\midrule
		\multicolumn{3}{l}{\textbf{Panel D: Model Diagnostics}} \\
		$R^2$ & 0.901 & 0.901 \\
		Adj.\ $R^2$ & 0.900 & 0.901 \\
		AIC & -28,720 & -29,030 \\
		BIC & -28,450 & -28,730 \\
		Residual skewness & 1.706 & 1.748 \\
		Residual kurtosis & 40.506 & 40.523 \\
		\bottomrule
		\multicolumn{3}{p{0.95\linewidth}}{\scriptsize
			\textbf{Note:} t-statistics in parentheses. $^{***}$, $^{**}$, and $^{*}$ denote significance at the 1\%, 5\%, and 10\% levels, respectively. HAC standard errors use 48 lags. All specifications include the full set of temporal fixed effects.}
	\end{tabular}
\end{table}

This section examines structural changes in the mapping from fundamentals and strategic conduct to clearing prices during the policy period. While the TPC mechanism was designed to truncate the upper tail of the price distribution, the associated regulatory environment may have also influenced bidding behavior. To test these dynamics, we estimate a regime-switching specification that allows behavioral elasticities to differ across the pre-policy and post-policy periods, benchmarking it against a constant-parameter baseline.

Table~\ref{tab:deterrence_results} shows that allowing behavioral elasticities to differ across regimes improves model fit, with the AIC declining from $-28{,}720$ to $-29{,}030$. The structural-break estimates point to a clear change in bid-to-price transmission during the post-policy window. In the pre-policy reference period of the break specification, the offer-ratio elasticity is positive and precisely estimated at 0.1167 (t = 9.46), consistent with an environment in which offer-side conditions can be reflected in clearing prices. In the post-policy period, the interaction term is negative and precisely estimated at -0.1405 (t = -7.53) and net elasty at implying that the overall sensitivity of prices to the offer ratio is largely neutralized and becomes close to zero, with a slightly negative implied net effect of -0.0238. This pattern is consistent with a substantial attenuation of markup pass-through under the post-policy regulation environment, rather than a definitive claim that strategic incentives or market power channels are eliminated.

A similar pattern is observed for scarcity pricing. Outside the policy window, prices respond sharply to tightening supply conditions, with SC elasticity -0.0985 and a t-statistic of -12.00. The policy interaction term is positive at 0.0673 with a t-statistic of 7.64, materially attenuating the pass-through of short-run reserve tightness. This is consistent with the cap compressing scarcity premia.

Finally, the regime indicator is negative at -0.0724 with a t-statistic of -7.68. This intercept shift and the interaction effects should be interpreted cautiously. While the model controls for observable fundamentals such as fuel costs and demand, the regime indicator captures the aggregate structural shift during the policy period. This likely reflects both the mechanical effect of the cap and the broader regulation context, including moral suasion and heightened scrutiny that may have discouraged aggressive bidding strategies independently of the hard cap. Nevertheless, the evidence supports the interpretation that the market operated under a fundamentally different structural regime post-policy, characterized by muted returns to strategic markups and reduced scarcity-driven price amplification.

\subsubsection{Strategic Avoidance and Price Bunching}
\label{sec:strategic_avoidance}

\begin{table}[ht]
	\centering
	\caption{Strategic Avoidance and Price Bunching Estimation Results}
	\label{tab:avoidance_results}
	\footnotesize
	\renewcommand{\arraystretch}{1.18}
	\setlength{\tabcolsep}{6pt}
	\begin{tabular}{lccc}
		\toprule
		\textbf{Dependent variable: $\ln P^{usep}_t$} & \textbf{5\%} & \textbf{10\%} & \textbf{15\%} \\
		\midrule
		
		\multicolumn{4}{l}{\textbf{Panel A: Trigger-Proximity region Dynamics}} \\
		\textbf{$D^{prox}_t$} & 0.0563 & 0.2015$^{**}$ & 0.1045\\
		& (0.50) & (2.56) & (1.84) \\
		\textbf{$M_t \times D^{prox}_t$} & 0.0056 & -0.0042$^{*}$ & -0.0007 \\
		& (1.05) & (-1.95) & (-0.74) \\
		\midrule
		
		\multicolumn{4}{l}{\textbf{Panel B: Core Fundamentals and Policy Controls}} \\
		\textbf{$\ln P^{lag}_{t}$} & 0.8082$^{***}$ & 0.8077$^{***}$ & 0.8072$^{***}$ \\
		& (87.23) & (86.54) & (86.38) \\
		\textbf{$\ln D_t$} & 0.7498$^{***}$ & 0.7505$^{***}$ & 0.7518$^{***}$ \\
		& (19.28) & (19.27) & (19.29) \\
		\textbf{$\ln F_{m,t}$} & 0.0786$^{*}$ & 0.0804$^{*}$ & 0.0836$^{*}$ \\
		& (2.08) & (2.12) & (2.20) \\
		\textbf{$\ln F^{lag}_{w,t}$} & 0.0531$^{*}$ & 0.0524$^{*}$ & 0.0542$^{*}$ \\
		& (2.21) & (2.18) & (2.28) \\
		\textbf{$\ln SC^*_t$} & -0.0229$^{***}$ & -0.0228$^{***}$ & -0.0217$^{***}$ \\
		& (-4.86) & (-4.92) & (-4.48) \\
		\textbf{$\ln OR^*_t$} & 0.1598$^{***}$ & 0.1610$^{***}$ & 0.1604$^{***}$ \\
		& (3.72) & (3.75) & (3.71) \\
		\textbf{$\ln VCQ_t$} & -0.3776$^{***}$ & -0.3771$^{***}$ & -0.3771$^{***}$ \\
		& (-14.42) & (-14.41) & (-14.39) \\
		\textbf{$\ln VCP_m$} & -0.0122 & -0.0120 & -0.0139 \\
		& (-0.28) & (-0.27) & (-0.31) \\
		\midrule
		
		\multicolumn{4}{l}{\textbf{Panel C: Model Diagnostics}} \\
		Adj.\ $R^2$ & 0.853 & 0.853 & 0.853 \\
		AIC & -17,040 & -17,040 & -17,050 \\
		Identified periods & 28 & 124 & 189 \\
		\bottomrule
		
		\multicolumn{4}{p{0.95\linewidth}}{\scriptsize
			\textbf{Note:} t-statistics in parentheses. $^{***}$, $^{**}$, $^{*}$ denote significance at the 1\%, 5\%, 10\%,, respectively. The estimation sample is restricted to the post-policy period. All specifications include the full set of temporal fixed effects.}
	\end{tabular}
\end{table}

The TPC mechanism places an explicit upper bound on the cumulative price trajectory. A key design concern is whether a cumulative trigger distorts incentives as the market approaches the threshold. In principle, the prospect of activating the cap could induce strategic avoidance, whereby generators restrain spot prices to preserve headroom and reduce the likelihood of a trigger. Alternatively, the trigger could become a salient reference point, leading to systematic threshold-related pricing dynamics as participants adjust bids in response to the remaining headroom. This section evaluates whether such proximity effects are detectable in the post-policy period.

To operationalize proximity, we construct the remaining margin as the difference between the administrative threshold and the lagged cumulative mix price. We then define a trigger-proximity indicator that equals one when the remaining margin falls within a specified percentage band of the threshold. The empirical design estimates a kinked response by interacting the trigger-proximity indicator with the continuous margin, allowing the slope of price formation to differ when the system enters the proximity window. This approach isolates whether price dynamics change discretely upon entering the proximity window and whether they exhibit systematic sensitivity to further reductions in headroom.

Table~\ref{tab:avoidance_results} reports results for three proximity definitions, 5\%, 10\%, and 15\%. Across specifications, the evidence does not support panic-induced braking. The level shift associated with entering the proximity window is not negative in any case, which is inconsistent with the prediction that prices collapse as the trigger approaches. In the 10\% specification, the trigger-proximity indicator is positive and statistically significant, while the margin interaction term is weakly significant. However, this pattern is not stable across alternative proximity definitions, and its magnitude is not sufficiently robust to characterize a pervasive threshold-driven strategy.

The sensitivity checks clarify the limits of inference. At the 5\% band, the number of identified proximity window periods is small, which reduces statistical power and increases sampling variability. At the 15\% band, the window is comparatively broad and likely includes periods in which the trigger is not yet behaviorally salient, attenuating any strategic effect that would be specific to tight headroom. The fact that the margin-related effect is not replicated across these alternative definitions suggests that any proximity response is, at most, episodic rather than systematic.

Overall, the three-threshold assessment yields no consistent evidence of strategic avoidance behavior or robust price bunching around the TPC trigger. Post-policy price formation appears to remain primarily anchored in fundamentals, with limited and non-robust sensitivity to the remaining headroom. From a market-design perspective, this indicates that the cumulative trigger has not generated economically significant distortions or systematic avoidance strategies.

\subsection{Synergistic Effects of the Dual-Defense Mechanism}

\subsubsection{Baseline Impact of Policy Mix on Average Price Levels}

\begin{table}[ht]
	\centering
	\caption{Policy Regime Effects on Average Spot Price Levels}
	\label{tab:policy_synergy_ols_log}
	\footnotesize
	\renewcommand{\arraystretch}{1.2}
	\setlength{\tabcolsep}{6pt}
	\begin{tabular}{lcc}
		\toprule
		\textbf{Dependent variable: $\ln P^{usep}_t$} & \textbf{Baseline} & \textbf{Policy-switch} \\
		\midrule
		\multicolumn{3}{l}{\textbf{Panel A: Policy Regime Variables}} \\
		\textbf{$R^{pol}_t$} &  & -0.0722$^{***}$ \\
		&  & (-7.067) \\
		\textbf{$R^{pol}_t \times \ln VCQ_t$} &  & -0.1954$^{***}$ \\
		&  & (-6.595) \\
		\textbf{$R^{pol}_t \times \ln VCP_m$} &  & -0.0051 \\
		&  & (-0.159) \\
		\midrule
		\multicolumn{3}{l}{\textbf{Panel B: VC Variables}} \\
		\textbf{$\ln VCQ_t$} & -0.1022$^{***}$ & 0.0043 \\
		& (-13.294) & (0.201) \\
		\textbf{$\ln VCP_m$} & -0.0561$^{***}$ & -0.0719$^{***}$ \\
		& (-4.627) & (-3.407) \\
		\midrule
		\multicolumn{3}{l}{\textbf{Panel C: Core Fundamentals}} \\
		\textbf{$\ln P^{lag}_{t}$} & 0.8565$^{***}$ & 0.8541$^{***}$ \\
		& (169.440) & (166.249) \\
		\textbf{$\ln D_t$} & 0.4325$^{***}$ & 0.5291$^{***}$ \\
		& (25.404) & (21.417) \\
		\textbf{$\ln F_{m,t}$} & -0.0232$^{*}$ & 0.0654$^{***}$ \\
		& (-2.338) & (4.594) \\
		\textbf{$\ln F^{lag}_{w,t}$} & 0.0646$^{***}$ & 0.0508$^{***}$ \\
		& (9.562) & (7.309) \\
		\textbf{$\ln SC^*_t$} & -0.0489$^{***}$ & -0.0443$^{***}$ \\
		& (-8.538) & (-8.214) \\
		\textbf{$\ln OR^*_t$} & 0.0756$^{***}$ & 0.0458$^{***}$ \\
		& (7.998) & (4.320) \\
		\midrule
		\multicolumn{3}{l}{\textbf{Panel D: Model Diagnostics}} \\
		Adj.\ $R^2$ & 0.900 & 0.901 \\
		AIC & -28,720 & -28,900 \\
		BIC & -28,450 & -28,610 \\
		\bottomrule
		\multicolumn{3}{p{0.95\linewidth}}{\scriptsize
			\textbf{Note:} t-statistics in parentheses. $^{***}$, $^{**}$, and $^{*}$ denote significance at the 1\%, 5\%, and 10\% levels, respectively. All specifications use HAC-robust standard errors and include month, year, and weekday fixed effects.}
	\end{tabular}
\end{table}

\begin{table*}[ht]
	\centering
	\caption{Quantile Regression Estimates Across the Price Distribution}
	\label{tab:quantile_results_usep_policy}
	\footnotesize
	\setlength{\tabcolsep}{6pt}
	\begin{tabular}{l p{2cm} p{2cm} p{2cm} p{2cm} p{2cm} p{2cm} }
		\toprule
		\textbf{Dependent variable: $\ln P^{usep}_t$} & \textbf{Q50} & \textbf{Q75} & \textbf{Q90} & \textbf{Q95} & \textbf{Q97.5} & \textbf{Q99} \\
		\midrule
		
		\multicolumn{7}{l}{\textbf{Panel A: VC Variables}} \\
		\textbf{$\ln VCQ_t$} 
		& -0.0052$^{**}$ & 0.0614$^{***}$ & 0.2622$^{***}$ & 0.3966$^{***}$ & 0.4900$^{***}$ & 0.4700$^{***}$ \\
		& (-2.764) & (6.529) & (11.787) & (10.763) & (8.159) & (3.920) \\
		\textbf{$\ln VCP_m$} 
		& -0.0112$^{***}$ & -0.0783$^{***}$ & -0.0813$^{***}$ & -0.0325 & 0.0202 & 0.0741 \\
		& (-5.988) & (-8.446) & (-3.792) & (-0.969) & (0.407) & (0.787) \\
		
		\midrule
		\multicolumn{7}{l}{\textbf{Panel B: Core Fundamentals}} \\
		\textbf{$\ln P^{lag}_{t}$} 
		& 0.9867$^{***}$ & 0.9653$^{***}$ & 0.9696$^{***}$ & 0.9984$^{***}$ & 1.0577$^{***}$ & 1.1884$^{***}$ \\
		& (4899.424) & (930.849) & (431.032) & (284.522) & (200.502) & (116.244) \\
		\textbf{$\ln D_t$} 
		& 0.0439$^{***}$ & 0.2508$^{***}$ & 0.5863$^{***}$ & 0.8066$^{***}$ & 1.0473$^{***}$ & 1.3096$^{***}$ \\
		& (25.219) & (28.655) & (28.122) & (23.326) & (18.858) & (12.107) \\
		\textbf{$\ln F_{m,t}$} 
		& 0.0122$^{***}$ & 0.0290$^{***}$ & 0.0409$^{**}$ & 0.0941$^{***}$ & 0.1248$^{***}$ & 0.2187$^{***}$ \\
		& (9.189) & (4.442) & (2.731) & (4.075) & (3.669) & (3.429) \\
		\textbf{$\ln F^{lag}_{w,t}$} 
		& 0.0016$^{***}$ & 0.0484$^{***}$ & 0.1249$^{***}$ & 0.1701$^{***}$ & 0.2007$^{***}$ & 0.3004$^{***}$ \\
		& (4.075) & (24.261) & (27.146) & (23.555) & (18.792) & (14.874) \\
		\textbf{$\ln SC^*_t$} 
		& -0.0031$^{***}$ & -0.0219$^{***}$ & -0.0664$^{***}$ & -0.1012$^{***}$ & -0.1423$^{***}$ & -0.1874$^{***}$ \\
		& (-10.944) & (-16.623) & (-22.754) & (-21.889) & (-21.044) & (-13.882) \\
		\textbf{$\ln OR^*_t$} 
		& 0.0014 & -0.0006 & 0.0493$^{***}$ & 0.1039$^{***}$ & 0.1855$^{***}$ & 0.1624$^{***}$ \\
		& (1.610) & (-0.153) & (5.270) & (7.298) & (8.968) & (4.196) \\
		
		\midrule
		\multicolumn{7}{l}{\textbf{Panel C: Policy Regime Interactions}} \\
		\textbf{$R^{pol}_t$} 
		& -0.0107$^{***}$ & -0.0429$^{***}$ & -0.0230$^{*}$ & 0.0024 & 0.0080 & -0.0160 \\
		& (-11.479) & (-9.475) & (-2.228) & (0.149) & (0.351) & (-0.370) \\
		\textbf{$R^{pol}_t \times \ln VCQ_t$} 
		& -0.0166$^{***}$ & -0.1724$^{***}$ & -0.4224$^{***}$ & -0.5814$^{***}$ & -0.6818$^{***}$ & -0.7275$^{***}$ \\
		& (-7.235) & (-14.668) & (-14.732) & (-12.226) & (-8.886) & (-4.771) \\
		\textbf{$R^{pol}_t \times \ln VCP_m$} 
		& 0.0075$^{*}$ & 0.0411$^{**}$ & -0.0604 & -0.2945$^{***}$ & -0.5001$^{***}$ & -0.7054$^{***}$ \\
		& (2.454) & (2.749) & (-1.764) & (-5.534) & (-6.384) & (-4.670) \\
		
		\midrule
		\multicolumn{7}{l}{\textbf{Panel D: Model Diagnostics}} \\
		Pseudo $R^2$ & 0.7964 & 0.7974 & 0.7601 & 0.7376 & 0.7248 & 0.6918 \\
		
		\bottomrule
		\multicolumn{7}{p{0.95\linewidth}}{\scriptsize
			\textbf{Note:} t-statistics are reported below coefficients. $^{***}$, $^{**}$, and $^{*}$ denote significance at the 1\%, 5\%, and 10\% levels, respectively. All specifications include month, year, and weekday fixed effects.}
	\end{tabular}
\end{table*}

Table~\ref{tab:policy_synergy_ols_log} reports OLS estimates under the log specification, testing whether the policy regime changes the way VCs map into spot prices. The results indicate a statistically and economically meaningful shift in the price-formation process under the dual-defense mechanism.

First, the policy regime is associated with a direct reduction in average spot prices. The coefficient on the regime indicator $R^{pol}_t$ is $-0.0722$ and statistically significant, implying a downward shift in the conditional mean of log prices after the policy is in force.

Second, the interaction terms point to a regime-contingent effect of VCQ. In the synergy specification, the standalone coefficient on $\ln VCQ_t$ is small and statistically indistinguishable from zero, suggesting that VCQ has limited explanatory power outside the policy regime within this specification. By contrast, the interaction term $R^{pol}_t \times \ln VCQ_t$ is negative at $-0.1954$ and highly significant, implying that the marginal effect of VCQ becomes substantially more negative during the policy regime. Equivalently, the implied policy-period semi-elasticity is $0.0043-0.1954=-0.1911$, indicating that higher VCQ is associated with materially lower spot prices once the policy regime applies.

Third, the VCP exhibits a comparatively stable association with spot prices across regimes. The coefficient on $\ln VCP_m$ remains negative and statistically significant, while the interaction term is economically small and statistically insignificant. This pattern is consistent with the VCP operating as a relatively regime-invariant anchor for price levels, in contrast to the quantity channel whose influence is conditional on the policy environment.

Overall, the OLS evidence is consistent with a complementary mechanism: the policy regime both lowers the conditional mean of prices directly and strengthens the price-suppressing association of VCQ, increasing the effectiveness of quantity-based hedging as a stabilizing force.

\subsubsection{Distributional Heterogeneity and Tail-Risk Mitigation}
\label{sec:qr_policy_interaction}

Table \ref{tab:quantile_results_usep_policy} reports quantile regression estimates under the log specification, highlighting how the policy regime reshapes price formation across the distribution. The results reveal a sharply asymmetric policy effect that intensifies toward the upper tail.

At the median, the policy dummy is negative and statistically significant, with a coefficient of −0.0107. The interaction between the policy regime and VCQ is also negative but economically modest at −0.0166, indicating that under normal market conditions the policy mainly operates as a mild downward shift in average prices rather than a strong binding constraint.

Moving into the upper quantiles, the nature of the policy effect changes markedly. At the 75th percentile, the interaction term with VCQ increases in magnitude to −0.1724, while at the 90th percentile it further deepens to −0.4224. This pattern strengthens at the extreme tail, where the interaction reaches −0.6818 at the 97.5th percentile and −0.7275 at the 99th percentile. These estimates imply that the marginal price impact of contract quantities is substantially amplified precisely when prices enter scarcity regimes.

A similar tail-dependent structure emerges for the interaction with VCP. While the coefficient is small and positive at the median, it turns strongly negative beyond the 95th percentile, reaching −0.5001 at the 97.5th percentile and −0.7054 at the 99th percentile. This reversal indicates that strike price signals lose their anchoring role in extreme conditions and instead reinforce the price-mitigating effect of the policy regime.

Importantly, the standalone policy dummy becomes statistically insignificant at the 95th percentile and beyond, while the interaction terms remain large and precisely estimated. This divergence confirms that the policy does not mechanically suppress prices across the distribution. Its primary effect is realized through strengthening the binding force of the vesting mechanism during high-price events.

Overall, the quantile results demonstrate that the policy regime operates as a state-contingent stabilizer. Its economic relevance is limited under normal conditions but becomes decisive in the upper tail, where the interaction with contract quantities and strike prices sharply reduces extreme price realizations.

\subsubsection{Volatility Dynamics under the Dual-Defense Policy Regime}
\label{sec:volatility_policy}

\begin{table}[ht]
	\centering
	\caption{Volatility Regression Results under the Dual-Defense Policy Regime}
	\label{tab:garch_policy_volatility_log}
	\footnotesize
	\renewcommand{\arraystretch}{1.15}
	\setlength{\tabcolsep}{6pt}
	\begin{tabular}{lcc}
		\toprule
		\textbf{Dependent variable: $\ln \hat{h}_t$} & \textbf{Baseline} & \textbf{Interaction} \\
		\midrule
		
		\multicolumn{3}{l}{\textbf{Panel A: VC Variables}} \\
		\textbf{$\ln VCQ_t$} & 0.0014 & 0.6363$^{***}$ \\
		& (0.0830) & (13.1250) \\
		\textbf{$\ln VCP_m$} & 0.0146 & -0.0617 \\
		& (0.5280) & (-1.3460) \\
		
		\midrule
		\multicolumn{3}{l}{\textbf{Panel B: Policy Regime Interactions}} \\
		\textbf{$R^{pol}_t$} &  & 1.7459$^{***}$ \\
		&  & (3.7440) \\
		\textbf{$R^{pol}_t \times \ln VCQ_t$} &  & -0.8490$^{***}$ \\
		&  & (-14.1040) \\
		\textbf{$R^{pol}_t \times \ln VCP_m$} &  & -0.0030 \\
		&  & (-0.0360) \\
		
		\midrule
		\multicolumn{3}{l}{\textbf{Panel C: Core Fundamentals}} \\
		\textbf{$\ln h^{lag}_{t}$} & 0.8264$^{***}$ & 0.8207$^{***}$ \\
		& (198.6540) & (191.7420) \\
		\textbf{$\ln P^{lag}_{t}$} & 0.3544$^{***}$ & 0.3588$^{***}$ \\
		& (23.0500) & (23.5320) \\
		\textbf{$\ln D_t$} & -0.0479 & 0.4266$^{***}$ \\
		& (-1.4250) & (9.1980) \\
		\textbf{$\ln F_{m,t}$} & -0.0151 & 0.0120 \\
		& (-0.7050) & (0.3340) \\
		\textbf{$\ln F^{lag}_{w,t}$} & -0.0068 & -0.0228$^{*}$ \\
		& (-0.5120) & (-1.6850) \\
		\textbf{$\ln SC^*_t$} & -0.0070 & 0.0001 \\
		& (-1.0240) & (0.0180) \\
		\textbf{$\ln OR^*_t$} & 0.0490$^{**}$ & 0.0674$^{***}$ \\
		& (2.4060) & (2.7570) \\
		
		\midrule
		\multicolumn{3}{l}{\textbf{Panel D: Model Diagnostics}} \\
		\textbf{$R^2$} & 0.9343 & 0.9349 \\
		\bottomrule
		
		\multicolumn{3}{p{0.95\linewidth}}{\scriptsize
			\textbf{Note:} t-statistics in parentheses. $^{***}$, $^{**}$, and $^{*}$ denote significance at the 1\%, 5\%, and 10\% levels, respectively. All specifications use HAC-robust standard errors with 48 lags and include the full set of temporal fixed effects (month, week, and year dummies).}
	\end{tabular}
\end{table}

This subsection investigates the impact of the dual-defense policy regime on second-moment price dynamics, measuring volatility through the logarithm of conditional variance. The analysis distinguishes between the baseline market structure and effects contingent on the policy regime.

The results presented in Table \ref{tab:garch_policy_volatility_log} demonstrate strong volatility persistence. The lagged log conditional variance serves as the dominant determinant and is precisely estimated, exhibiting a coefficient of approximately 0.82 across specifications. Similarly, the lagged price term remains statistically significant with a magnitude of approximately 0.35. These findings confirm the presence of pronounced volatility clustering, where historical instability serves as a strong predictor of subsequent variance.

While the baseline specification yields no systematic association between VCQ and volatility, a distinct structural pattern emerges once policy timing is explicitly incorporated. In the pre-policy period, VCQ exerts a positive and highly significant influence on volatility, a finding that supports the liquidity-squeeze hypothesis. Conversely, VCP lacks statistical significance, indicating the absence of a robust linkage between the vesting price and second-moment dynamics within the current specifications.

The interaction specification reveals a structural reversal during the post-policy period. The policy regime indicator is positive and significant, reflecting the elevated baseline stress that characterized the market following policy implementation. However, the interaction term between the policy regime and VCQ is strongly negative and highly significant, dominating the pre-policy VCQ elasticity. Consequently, the implied net effect of VCQ becomes negative in the post-policy period. This suggests that the dual-defense mechanism effectively neutralizes the volatility penalty typically associated with high VCQ levels, thereby mitigating the liquidity-withdrawal channel.

Market fundamentals also exhibit state-dependent associations with volatility. System demand shows a positive and significant relationship in the post-policy specification. Furthermore, the offer ratio remains a positive and significant driver, implying that strategic bidding behaviors continue to contribute to price instability even after policy implementation. The model demonstrates strong explanatory power, attaining a coefficient of determination of approximately 0.935.

In summary, the interaction results imply a regime-dependent stabilization in volatility transmission rather than a uniform reduction in variance levels. High VCQ acts as a destabilizing force in the pre-policy period, whereas the post-policy regime effectively reverses this relationship. Thus, the dual-defense mechanism succeeds in suppressing price levels while simultaneously mitigating volatility amplification through the VCQ channel.

\section{Conclusion and Policy Recommendations}

This study empirically evaluates the efficacy of the dual-defense mechanism in Singapore which integrates structural VC with a TPC mechanism. Utilizing a high-frequency dataset from 2021 to 2024, we reconstructed latent supply-side dynamics through MFMSSA and employed a suite of econometric strategies including regime-switching OLS, structural break models, kinked regression designs, and GARCH models to quantify the impact of these regulations on price levels, tail risks, and volatility dynamics.

\subsection{Summary of Empirical Findings}

Our analysis yields three primary conclusions regarding the mechanics of market regulation.

First, the VC mechanism exhibits a structural trade-off between price suppression and volatility management. The empirical results demonstrate that while the VCQ effectively lowers wholesale prices, its dampening effect amplifies by approximately 7.8 times from a baseline elasticity of -0.0591 in normal regimes to -0.4617 during scarcity. However, this price-suppressing role comes with a significant state-dependent stability cost. The volatility analysis reveals that high VCQ exacerbates market instability precisely when the market is tight. While VCQ exerts a mild stabilizing effect with an elasticity of -0.0423 in normal regimes, this relationship reverses sharply during scarcity conditions, where the net volatility elasticity rises to 0.5649. This confirms that static quantity hedging induces a liquidity squeeze in the residual spot market during threshold-breaching episodes, which lowers the price floor but increases the variance of price innovations. Meanwhile, the two components of the vesting framework demonstrate distinct functional specializations across the price distribution. The VCP serves as a weak constraint under typical conditions, with a small median elasticity of -0.0071, but becomes the dominant anchor for tail risks, where the quantile estimates indicate a potent elasticity of -0.2430 at the 99th percentile. Conversely, VCQ continues to dampen high prices across the upper tail, with an elasticity of -0.0710 at the 95th percentile and -0.0811 at the 99th percentile. This implies that VCQ mandates suppress the conditional mean and broad scarcity rents, while the VCP mechanism is required to truncate extreme tail realizations.

Second, the 2023 policy adjustment coincides with a structural shift in bidding behavior, effectively mitigating the influence of strategic bidding on price formation. The structural break analysis reveals a decoupling of the relationship between offer behaviors and clearing prices, suggesting that the potential exercise of market power has been constrained. Prior to the 2023 policy adjustment, prices exhibit a positive offer ratio elasticity of 0.1167, consistent with substantial scope for markups to be transmitted into clearing outcomes; under the combined VC and TPC regime, this sensitivity is largely neutralized, with the implied net effect close to zero at -0.0238. Further analyzing the specific design of the TPC cumulative trigger, we find no evidence of panic-induced braking or systematic avoidance. In the 10 percent proximity window, the intercept shifts positively by 0.2015 while the margin interaction term is weakly negative at -0.0042. This pattern implies that generators tended to price up to the limit rather than artificially depressing bids to preserve headroom, indicating that the cumulative cap did not induce pre-emptive distortions.

Third, the dual-defense mechanism exhibits a distinct functional synergy between price level suppression and volatility management. This policy complementarity is most pronounced in the context of tail-risk mitigation, where the interaction coefficient between the policy regime and VCQ at the 99th percentile reaches -0.7275, a magnitude that far exceeds the baseline effect. Furthermore, the volatility analysis indicates that the regulation effectively resolved the liquidity trade-off associated with high VCQ. Under the dual-defense mechanism, the structural sensitivity of volatility to contract quantity is structurally reversed. While the baseline elasticity of VCQ suggests a significant destabilizing effect of 0.6363 outside the policy window, the policy interaction term of -0.8490 provides a substantial offset. Consequently, the net elasticity of −0.2127 implies that the dual-defense mechanism breaks the link between high VCQ and spot market instability, allowing price suppression without triggering additional volatility.

\subsection{Policy Recommendations}

Based on these empirical findings, we propose targeted recommendations for refining the market design.

We affirm the necessity of the TPC mechanism but note its inherent limitations. Since the MAP is calculated as a 24-hour rolling average, the current trigger method is relatively simple and reactive. This structural feature may allow generators to engage in tacit collusion by exercising market power to keep the MAP just below the trigger zone, thereby evading the cap while maintaining high prices. Furthermore, in scenarios of genuine supply shortage, a rigid cap might unduly restrict the legitimate revenue recovery of generators.

To address these issues, we recommend introducing a monitoring mechanism specifically targeting the critical proximity range of the MAPT. As generators may strategically control the MAP within this buffer zone to avoid triggering the TPC, enhanced surveillance of bidding behaviors in this marginal window is essential to detect and mitigate potential market power abuse. Additionally, the regulator should consider dynamically adjusting the MAPT based on the potential market state reflected by the RUSEP. This linkage would prevent the price cap from becoming overly restrictive during genuine cost shocks while preserving its function as a circuit breaker against speculative spikes.

Regarding the VC mechanism, it plays a vital role given the insufficient liquidity in the forward market where participants lack access to transparent and fair price signals. The VC serves as a necessary guidance tool to anchor expectations. However, excessive reliance on this mechanism should be avoided. Although empirical results confirm its ability to suppress average prices, the analysis also indicates that high VCQ amplifies market volatility. Moreover, market outcomes are highly sensitive to the VCP parameter. If the price is set inappropriately, it may inadvertently strengthen the incentive for generators to exercise market power. Therefore, the calibration of VC parameters requires caution to balance stability with market fluidity.


\subsection{Limitations and Future Research}

This study provides an econometric assessment of key regulation policies in Singapore’s electricity market, but several limitations remain and motivate future research.

First, data constraints and limited transparency restrict the scope of inference. Singapore has low liquidity in the standardized forward market, while a substantial share of trading occurs through private bilateral contracts. This opacity prevents a full characterization of strategic interactions and contracting positions across participants. As a result, our analysis cannot identify the complete impact of forward contracting on wholesale prices. The VC information used here covers only the government-mandated component with administered quantities and prices, and therefore may not capture the price effects arising from the broader unobserved forward market.

Second, the empirical design emphasizes the direct effects of the policy variables included in the specification. Some micro-level determinants of price formation are not observable in the available data and are therefore omitted. Future research can extend the analysis by incorporating additional drivers of price outcomes and by modelling heterogeneity in firm behavior and its implications for market equilibrium.

	\bibliographystyle{elsarticle-harv}  
	\bibliography{mp_ref.bib}
\end{document}